\documentclass[epj]{svjour}

\usepackage{xparse}
\usepackage{amsmath}
\usepackage{graphicx}
\usepackage{physics}
\usepackage{hyperref}
\usepackage{bm}
\usepackage{float}
\usepackage{xcolor}

\usepackage[utf8]{inputenc}
\usepackage[T1]{fontenc}

\usepackage[switch]{lineno}

\newcommand{\rr}[1]{\textcolor{black}{#1}}
\newcommand{\ra}[1]{\textcolor{black}{#1}}

\begin{document}


\title{Estimating theoretical uncertainties of the two-nucleon observables
by using backpropagation}

\author{K. Topolnicki \inst{1} \and R. Skibiński \inst{1} \and J. Golak \inst{1}}

\institute{$^{1}$ M. Smoluchowski Institute of Physics, Jagiellonian University, Profesora Stanisława \rr{Łojasiewicza} 11, Kraków, PL-30348, Poland}

\date{Received: date / Revised version: date}

\abstract{
    We present a novel approach to calculating theoretical uncertainties in 
	few-nucleon calculations, making use of automatic differentiation via 
	backpropagation, which is particularly efficient when 
	there are many input variables but only a few outputs.
	\ra{The methods described in this paper constitute tools that can be used to investigate
	the properties of scalar functions used to define nuclear potentials and quantify their contribution to the
	uncertainty of few nucleon calculations.} 
	We demonstrate 
	\rr{these methods} 
	in deuteron bound state and nucleon - nucleon scattering calculations. 
	Backpropagation, implemented in the \textbf{Python}
	\textbf{pytorch} library, is used to calculate the gradients with respect to
	model parameters and propagate errors from these parameters to the deuteron binding energy and selected phase-shift parameters. The uncertainty
	values obtained using this approach are validated by directly sampling from the
	potential parameters. {We find very good agreement between two ways of estimating that uncertainty. }
	\PACS{
		{03.65.Ge}{Bound states, quantum mechanics} \and
		{24.10.-i}{Nuclear reactions, models of}
     } 
} 

\maketitle

\section{Introduction}

The progress of research on the nuclear potential, and in particular the increasing precision of scattering experiments, has drawn researchers' attention to estimating the uncertainty of the theoretical models used~\cite{Dobaczewski_2014}. In the first stage, the influence of experimental uncertainties on the determination of free potential parameters was examined. An important step in that direction was the preparation of a self-consistent database for nucleon-nucleon scattering data by the group from Granada~\cite{Navarro_GranadaBase} and using it to determine the uncertainty of the parameters of the OPE-Gausian~\cite{OPEGaussian} and other potentials along with the corresponding correlation coefficients. Also the introduction of new generations of few-nucleon forces from the Chiral Effective Field Theory, see e.g. reviews~\ra{\cite{RevModPhys.81.1773,Epelbaum_review2012,RevModPhys.92.025004}}, 
raises the important question of determining the errors associated
with few-nucleon quantum mechanical calculations. The concept of determining the correlations between potential parameters was taken over by the Bochum group, which delivered uncertainties in the potential parameters and the resulting uncertainties in nucleon-nucleon phase shifts for the chiral SMS potential~\cite{Reinert_SMS}. Using statistical analysis methods in~\cite{Skibinski_uncer2018,Volkotrub_2020} it was examined how these uncertainties are transferred to observables in elastic nucleon-deuteron scattering and in the deuteron breakup reaction. These methods involve calculating observables for various sets of potential parameters and then examining the obtained probability distributions for individual observables. These distributions depend on both the reaction energy and kinematic variables, e.g. scattering angles. It turned out that the uncertainties of the observables related to the uncertainties of the potential parameters are relatively small and usually remain below 1\% for the differential cross section. Similar study but based on approximated solution of three-nucleon problem has been presented in~\cite{Miller_2023}, where also other types of uncertainties are discussed. In~\cite{Witala-emu3} some uncertainty related to 3NF parameters have been estimated for the nucleon-deuteron scattering by using the approximated solutions of the Faddeev equations. 
\ra{The rapidly increasing computational 
complexity with the number of nucleons makes investigations based on statistical 
methods very challenging for many-nucleon systems. Recently new technique based on reduced-basis method
has been developed to create and train emulators for many-nucleon systems, see \cite{RevModPhys.96.031002,arxivNonHermitianQuantum}.}

It should be remembered that among other types of theoretical uncertainties related to the chiral potential two play a dominant role. The first is the uncertainty related to the value of the \ra{cut-off parameter} used in the potential regularization. The second source of uncertainty is the neglect of higher terms of chiral expansion in the nuclear potential, which leads to the so-called truncation errors. Estimating the uncertainties associated with different cut-off values involves repeating the calculations, in practice for a few values (four in case of the SMS chiral potential~\cite{Reinert_SMS}). It should be remembered that the regulator value also affects the values of the free potential parameters, so in fact we are dealing with several versions of the potential.
The uncertainty associated with using different cut-off values is typically of the order of 1-3\% for nucleon-deuteron scattering and for deuteron and $^3$He photodisintegration~\cite{Urbanevych}. The estimation of truncation errors was proposed in~\cite{SCS1,SCS2}, which treated observables as quantities subject to chiral expansion. This recipe was then extended to many-nucleon processes in~\cite{Epelbaum_Towards}. The truncation errors remain typically below 10\% depending strongly on the order of chiral expansion and reaction energy for nucleon-deuteron scattering and for $^2$H or $^3$He photodisintegration~\cite{Urbanevych}.
It was shown that this prescription is consistent with the estimation of truncation errors resulting from Bayesian analysis methods \ra{with a particular choice of priors}~\cite{Furnstahl:2015rha,Melendez:2019izc}. The role of various uncertainties in the nuclear structure calculations as well as various aspects of uncertainty quantification are currently the subject of many efforts, see e.g.~\cite{Maris2022,Millican2024,Becker2024} and references therein. 

Despite the fact, that both the cut-off dependence and truncation errors of two- and many-nucleon observables turned out to be usually bigger then the above mentioned dependence on potential parameters, in this 
work we focus on the uncertainty of two-nucleon observables arising from uncertainty of potential parameters. 
We demonstrate that software libraries designed for machine learning can be used to investigate
these uncertainties. Using these tools to propagate errors from model parameters
to observables is, to the best of our knowledge, a novel alternative to using 
standard statistical methods. 

\ra{Few nucleon potentials are typically expressed as a linear combination of isospin - spin - momentum operators
and scalar functions. The form of the operators is fixed and the scalar functions effectively define the force.
In practice, the possibility to investigate the properties of scalar functions that define nuclear interactions
and quantify their contribution to the uncertainty of few nucleon calculations using the tools described in this paper
is valuable in the development and debugging of 
a numerical implementation of these potentials.} 

\rr{We} use the \textbf{Python} library \textbf{pytorch} \cite{NEURIPS2019_9015} to implement the numerical calculation of the deuteron bound state energy
and nucleon-nucleon scattering observables. 
This library allows for an efficient implementation of numerical calculations. 
The numerical operations within this library are vectorized 
making it is possible
to automatically parallelize their execution.
Most importantly, \textbf{pytorch} offers the 
possibility to use backpropagation \cite{JMLR}
to calculate the gradients 
of the results obtained from computations. This will be used to 
estimate errors of the bound state energy and nucleon - nucleon phase shifts 
that result from uncertainties in
the two-nucleon (2N) potential.

\ra{The theoretical introduction to the calculations is formulated in the
language of ``tensors". This term is not used to refer to the mathematical
construct but to multidimensional arrays. 
Numerical operations
mapped over multidimensional arrays are relatively straightforward to parallelize
over multiple CPU threads on graphics processing units making ``tensors"
a popular fundamental data type in many python libraries designed for machine learning and numerical calculations.
Formulating the calculations in this language makes them easier to translate
to other libraries, for example \textbf{tensorflow} \cite{tensorflow2015-whitepaper}, or if automatic
differentiation is not required \textbf{numpy} \cite{harris2020array}. 
} 

\ra{The input to our calculation is a set of scalar functions that define the
two nucleon potential. 
For our purposes we will be using the well known Bonn~B \cite{Machleidt1989} potential
but by simply changing the scalar functions, other more modern potentials can be used.
Additionally, we will assume that the values of these scalar functions are normally distributed
to simulate errors associated with the functions. The parameters of these distributions are 
an additional part of the input to our algorithms. 
Since the scalar functions are represented as a discrete lattice of values, 
the derivatives of the calculated observables with respect to the, hundreds of thousands,
of these values needs to be calculated in order to investigate the influence of errors on the input
on the calculated quantities.
For this reason the calculations in this paper rely heavily on efficient algorithms to perform automatic
differentiation. 
Many current libraries including \textbf{pytorch} rely on, so called, backward mode automatic differentiation
or backpropagation. In this approach the algorithm keeps track of the numerical operations used to calculate
a given result in the form of a graph. When the calculation is complete, the graph can be followed backwards
from the output to the inputs and the chain rule can be used to calculate the relevant gradients. Nodes of the
graph may point to intermediate results of the calculation making this process more efficient.
In another approach to automatic differentiation, forward mode automatic differentiation, the
calculation is performed in only one direction. This type of procedure is also widely used, for example in the study \cite{PhysRevX.6.011019}.
In a very basic form of forward mode automatic differentiation the numerical types used in the calculation
can be replaced by appropriate dual numbers \cite{Angeles1998}. The additional information contained in these numbers
makes it possible to track a given derivative throughout the computation.
} 

The paper is organized as follows. In Section \ref{formalism} we outline our approach to
calculating the deuteron bound state. Subsection \ref{implementation} contains a 
description of the implementation of the calculation in the \textbf{pytorch}
library and Subsection \ref{res} contains a discussion of the numerical results
and obtained error estimates.
Next, in Section \ref{nnintroduction} we outline the theoretical
formalism used in nucleon - nucleon scattering calculations.
Subsection \ref{nntorch} contains details related to the implementation
of these calculations in the \textbf{pytorch} library and
subsection \ref{nnresults} contains a discussion of the numerical results
and obtained error estimates.
Finally, Section \ref{so}
contains the summary and outlook.

\section{The deuteron bound state}
\label{formalism}

We will use the Schr{\"o}dinger equation in integral form to 
calculate the deuteron bound state:
\begin{equation}
	\left( E_{d} - H_{0} \right)^{-1} V \ket{\phi_{d}} \equiv K(E_{d}) \ket{\phi_{d}} = 
	\ket{\phi_{d}}.
	\label{de}
\end{equation}
In this equation $\ket{\phi_{d}}$ is the deuteron bound state, $E_{d}$ is 
the deuteron binding
energy, $H_{0} = \frac{\vb{p}^{2}}{m}$ is the kinetic energy operator
with $\vb{p}$ being the relative 2N momentum operator, $V$ is the
2N potential operator and $m$ is the nucleon mass. For practical
calculations 
the deuteron binding energy $E_{d}$ in \eqref{de} is replaced by $E$ and a slightly 
different equation is solved:
\begin{equation}
	K(E) \ket{\phi} = \lambda \ket{\phi}\;.
	\label{de1}
\end{equation}
The bound state is found by searching for a
value of $E$ such that there exists an eigenvalue $\lambda$ in \eqref{de1} 
equal to
$1$ up to assumed precision. This value of $E$, if found, is taken as the deuteron
bound state energy and the corresponding eigenstate $\ket{\phi}$ 
as the deuteron bound state.

The fundamental element in this calculation is the 2N potential
$V$. Following \cite{Golak2009} 
we will assume that the potential satisfies:
\begin{equation}
	\mel{t' m_{t}'}{V}{t m_{t}} = \delta_{t' t} \delta_{m_{t}' m_{t}}
	V^{t m_{t}}
	\label{2npot}
\end{equation}
where $t$ ($t'$) is the total 2N isospin in the initial (final) state,
$m_{t}$ ($m_{t}$') is the projection of the isospin in the initial
(final) state and $V^{t m_{t}}$ is the potential operator in the
joined spin and momentum space of the 2N system. 

The 2N potential can be expanded into a linear combination of scalar
functions and operators \cite{Wolfenstein1954}:
\begin{equation} 
	\mel{\vb{p}'}{V^{t m_{t}}}{\vb{p}} = \sum_{i = 1}^{6} 
	v^{t m_{t}}_{i}(p' , p , \vu{p}' \cdot \vu{p}) w_{i}(\vb{p}' , \vb{p}) \label{com}
\end{equation}
where $v^{t m_{t}}_{i}(p' , p , \vu{p}' \cdot \vu{p})$ are scalar functions of
the relative 2N momentum magnitudes $|\vb{p}| = p$ ($|\vb{p}'| = p'$) in the initial (final) 
state and $w_{i}(\vb{p}' , \vb{p})$ are operators in
the spin space of the 2N system \cite{Golak2009}.
Since in this section we are concentrating on deuteron bound state calculations, we will be
using only the $V^{0 0}$ part of the potential
and drop the upper index $^{0 0}$ in $V^{0 0}$ and $v^{0 0}_{i}$ for brevity. 

Our choice for the set of the $w_{i}$ operators is the same as in \cite{Golak2009}:
\begin{align} \nonumber
w_{1}(\vb{p}' , \vb{p}) = 1, \\ \nonumber
w_{2}(\vb{p}' , \vb{p}) = \bm{\sigma}(1) \cdot \bm{\sigma}(2), \\ \nonumber
w_{3}(\vb{p}' , \vb{p}) = i (\bm{\sigma}(1) + \bm{\sigma}(2)) \cdot (\vb{p} \cross \vb{p}'), \\ \nonumber
w_{4}(\vb{p}' , \vb{p}) = \bm{\sigma}(1) \cdot (\vb{p} \cross \vb{p}')  \bm{\sigma}(2) \cdot (\vb{p} \cross \vb{p}'), \\ \nonumber
w_{5}(\vb{p}' , \vb{p}) = \bm{\sigma}(1) \cdot (\vb{p}' + \vb{p})  \bm{\sigma}(2) \cdot (\vb{p}' + \vb{p}), \\ 
	w_{6}(\vb{p}' , \vb{p}) = \bm{\sigma}(1) \cdot (\vb{p}' - \vb{p})  \bm{\sigma}(2) \cdot (\vb{p}' - \vb{p}) ,\label{al}
\end{align}
where $\bm{\sigma}(i)$ is a vector of Pauli matrices acting in the space
of particle $i$. The 2N potential is
determined by the scalar functions $v_{i}$.
These scalar functions will be represented numerically as discrete
values on a lattice. More specifically, if the momenta $p'$, $p$ 
share the same discrete values:
\begin{equation}
	p_{1} , p_{2} \ldots p_{N} \label{p}
\end{equation}
and the cosine of the angle between the momenta, $\vu{p}' \cdot \vu{p}$, is also discretized and can take
on values:
\begin{equation}
	x_{1} , x_{2} \ldots x_{M} ,\label{x}
\end{equation}
then the scalar functions will be represented as an array $\mathcal{V}$
whose elements are:
\begin{align} \nonumber
	&\mathcal{V}_{ijkl} = v_{i}(p_{j} , p_{k} , x_{l}) ,\\ \nonumber
	&i = 1 \ldots 6 \,\, , \,\, j = 1 \ldots N ,\\ 
	&k = 1 \ldots N \,\, , \,\, l = 1 \ldots M . \label{vcal} 
\end{align}
When integration over momentum or angle is required, integration weights
$\mathcal{W}^{p}_{i = 1 \ldots N}$ or $\mathcal{W}^{x}_{i = 1 \ldots M}$
will be used for $p$, $p'$ or $x$ respectively. Our calculations make
use of Gaussian quadrature points and weights.

Our numerical calculation will use the array \eqref{vcal} as input.
In the first step, the machine learning library \textbf{pytorch}
\cite{NEURIPS2019_9015}
is used to
calculate the partial wave decomposition of the 2N potential 
which in turn will be used
to solve the eigenproblem \eqref{de1}. Finally, we will use the backpropagation
algorithm implemented in \textbf{pytorch} to calculate gradients of
the resulting eigenvalues $\lambda$ from \eqref{de1} with respect to the elements of $\mathcal{V}$. This will
allow us to propagate the theoretical uncertainties from the
$\mathcal{V}$ array
to the eigenvalues $\lambda$ that are closest to $1$ for a given energy. 
Since the dependence of the eigenvalue closest to $1$ on the energy $\lambda(E)$
is used to estimate the deuteron binding energy $\lambda(E_{d}) = 1$,
knowing the uncertainty of $\lambda(E)$ will also determine the
uncertainty of the deuteron binding energy estimate.

\subsection{Implementation using \textbf{pytorch}}
\label{implementation}

The partial wave decomposition of the 2N potential was performed using
the procedure from \cite{Golak2009}. Since our goal is to use
backpropagation, the implementation was written
entirely using methods and functions from the \textbf{pytorch} library. 

Due to the rotational invariance of the 2N potential, its partial-wave matrix element is determined in \cite{Golak2009} by a simpler function $H$:
\begin{equation}
	\mel{p' ( l' s ) j m_{j}}{V}{p ( l s ) j m_{j}} = H(p',p;l',l,s,j). \label{elmt}
\end{equation}
Here 
\begin{equation}
	\ket{p(ls)jm_{j}}
	\label{pwstates}
\end{equation}
are states with the magnitude of the relative momentum $p$. The states carry information about the orbital angular momentum
of the two-particle system, $l$, which is coupled with the spin $s$ to form 
a state with the total angular momentum $j$ with projection $m_{j}$.
Practical numerical calculations are carried out by considering only a finite subset of
partial wave states $\ket{p(ls)jm_{j}}$. For deuteron calculations it is sufficient
to consider only two partial wave states, however in our calculations
we consider all partial wave states with $j \le 5$ for bench-marking purposes. 

The explicit form of $H$, for arguments where it has non zero values, is \cite{Golak2009}:
\begin{align} \nonumber
	H(p',p;l',l,s,j) = 8 \pi^{2} \int_{-1}^{1} \text{d}(\cos{\theta'})
	\frac{1}{2 j + 1} \sum_{m_{j} = -j}^{j} \\ \nonumber
	\sum_{m_{l}' = -l'}^{l'} c(l',s,j;m_{l}',m_{j} - m_{l}',m_{j}) \\ \nonumber
	\sum_{m_{l} = -l}^{l} c(l,s,j;m_{l},m_{j} - m_{l},m_{j}) \\ \nonumber
	Y_{l'm_{l}'}^{*}(\theta' , 0) Y_{lm_{l}}(0 , 0) \\
	\mel{\vb{p'};s,m_{j}-m_{l}'}{V}{\vb{p};s,m_{j}-m_{l}} \label{h}
\end{align}
where $\vb{p} = (0 , 0 , p)$, $\vb{p}' = p' (\sin{\theta'} , 0 , \cos{\theta'})$, \newline $c(j_1, j_2, j; m_1, m_2, m)$ are Clebsh-Gordan coefficients,
\newline $Y_{l m} \left(\theta, \phi \right)$ are spherical harmonics and
$\ket{\vb{k};s,m_{s}}$ are states with relative 2N momentum $\vb{k}$ and spin $s$
with projection $m_{s}$.  
The numerical calculation of \eqref{elmt} can be 
obtained by substituting \eqref{com} in \eqref{h}. Assuming that the
momentum magnitudes and cosines of angles are discretized, \eqref{elmt} can be approximated by:
\begin{equation}
	\mel{p_{j} \alpha}{V}{p_{k} \beta} \approx \sum_{i = 1}^{6} \sum_{l = 1}^{M} 
	\mathcal{H}_{i j \alpha k \beta l} 
	\mathcal{W}^{x}_{l} \mathcal{V}_{ijkl} \equiv \mathcal{V}^{pwd}_{j \alpha k \beta} , \label{vpwd}
\end{equation}
where $\mathcal{W}^{x}_{l}$ are integration weights for \eqref{x}, 
and $\alpha$ ($\beta$) are the complete set of discrete quantum numbers
in the final (initial) state from \eqref{elmt}. Finally, the values of $\mathcal{H}_{i \alpha j \beta k l}$
are calculated by taking the integrand from \eqref{h} with $p' = p_{j}$, $p = p_{k}$ and substituting the $w_{i}$ operator from \eqref{al} for the potential $V$.

Each of the three arrays $\mathcal{H}$, $\mathcal{W}^{x}$ and $\mathcal{V}$ is represented
by \textbf{pytorch} ``tensors''. 
The first ``tensor'' $\mathcal{H}$ was calculated, for a given set of discrete quantum numbers,
using symbolic programming in \textit{Mathematica} 
\cite{Mathematica}. The resulting expressions were translated into \textbf{Fortran} and then 
used to create a \textbf{Python} module with \href{https://numpy.org/doc/stable/f2py/}{\textbf{f2py}},
a part of the \textbf{numpy} library \cite{harris2020array}.
The second ``tensor" $\mathcal{W}^{x}$ contains Gaussian integration weights.
Only the last ``tensor" $\mathcal{V}$ requires the gradient
since values of its elements are assumed to have uncertainties.
The values of $\mathcal{V}$ are calculated in a separate program for the Bonn~B \cite{Machleidt1989} potential.

The sum in \eqref{vpwd} is calculated using the \textbf{einsum} method from the \textbf{pytorch}
library. The created $\mathcal{V}^{pwd}$ ``tensor'' together with the integration weights for \eqref{p}, $\mathcal{W}^{p}_{l}$,
can be used to construct a discrete representation of the $K(E)$ operator from \eqref{de1}. The action of this operator
on a 2N state $\ket{\phi}$ given in partial wave representation:
\begin{equation}
	\mel{p_{j} \alpha}{K(E)}{\phi} \approx \sum_{k = 1}^{N} \sum_{\beta} \mathcal{K}(E)_{j \alpha k \beta} \braket{p_{k} \beta}{\phi},
\end{equation}
where the second sum is over a finite subset of all possible discrete quantum numbers $\beta$,
is given in terms of the $\mathcal{K}(E)_{j \alpha k \beta}$ array:
\begin{align} \nonumber
	\mathcal{K}(E)_{j \alpha k \beta} = \\ \nonumber
	\left( E - \frac{p_{j}^{2}}{m} \right)^{-1} \mel{p_{j} \alpha}{V}{p_{k} \beta} p_{k}^{2} \mathcal{W}^{p}_{k} \equiv \\
	\left( E - \frac{p_{j}^{2}}{m} \right)^{-1} \mathcal{V}^{pwd}_{j \alpha k \beta} p_{k}^{2} \mathcal{W}^{p}_{k}.
	\label{kdeuteron}
\end{align}
The $\mathcal{K}(E)_{j \alpha k \beta}$ ``tensor'' can be constructed 
using \textbf{pytorch} methods and then reshaped into a two dimensional matrix.
This matrix can be used to solve 
the eigenproblem \eqref{de1} with the \textbf{eigvals} function. 
Because all operations, including the calculation of the eigenvalue, are performed
within the \textbf{pytorch} library it is possible to calculate the gradient
of a specific eigenvalue with respect to elements of $\mathcal{V}$:
\begin{equation}
\frac{\partial{\lambda}}{\partial{\mathcal{V}_{ijkl}}}
\end{equation}
using backpropagation. 
In practice this is done in two steps. First the \textbf{backward()} method is called on a given eigenvalue. Next the gradient can be read from the \textbf{grad} field of the \textbf{pytorch} ``tensor'' corresponding to $\mathcal{V}$. 
In practice we will calculate gradients of eigenvalues that are closest to $1$ for a given energy $E$.

Finally, we introduce uncertainties to the elements of $\mathcal{V}$
assuming their standard deviations arise from the uncertainties of the scalar
functions $v$.
For the purposes of this paper we use the following choice:
\begin{equation}
	\sigma(\mathcal{V}_{ijkl}) = \alpha(i) \abs{\mathcal{V}_{ijkl}}, \label{paramerrors}
\end{equation}
in which the standard deviation $\sigma$ is not constant across the scalar function values but
proportional to their absolute value. We allow the proportionality factor $\alpha(i)$
to have different values for different operators $w_{i}$ from \eqref{com}.
\ra{The values of $\alpha(i)$ are an important part of the input to our calculations,
changing them allows the investigation of the influence of uncertainties associated with 
scalar functions on the uncertainty associated with the the calculated observables.} 
Additionally we assume that there are no correlations, so that the errors can be
propagated using:
\begin{equation}
\sigma(\lambda) = \sqrt{\sum_{ijkl} \sigma(\mathcal{V}_{ijkl})^{2} 
	\left(\frac{\partial{\lambda}}{\partial{\mathcal{V}_{ijkl}}}\right)^{2}}. \label{err}
\end{equation}
Adding correlations is possible but it \ra{would significantly increase the memory requirements} of the calculation
\ra{since we need to treat the hundreds of thousands of values 
in $\mathcal{V}$ as variables for automatic differentiation}
; we plan to investigate correlated errors in future work. Note that we assign errors to individual elements of $\mathcal{V}$. 
The scalar functions are continuous but we associate uncertainties only with the values of the scalar functions
at specific points.

The next section contains numerical results and describes the procedure to obtain the deuteron binding energy
and its uncertainty.

\subsection{Results for the deuteron binding energy}
\label{res}

\begin{figure*}[h]
	\begin{center}
		\includegraphics[width = 0.75 \linewidth]{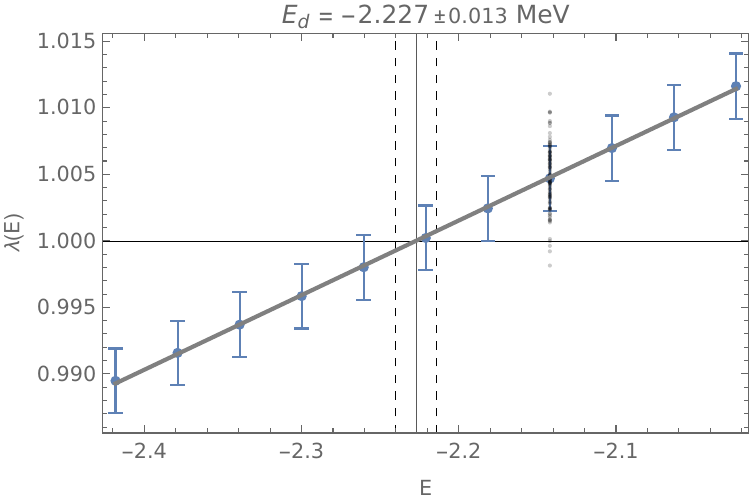}
	\end{center}
	\caption{Eigenvalues closest to $1$ for different
	values of $E$ in MeV from \eqref{de1}. 
	The standard deviation of the scalar function values is 
	\ra{taken to be} $\sigma(\mathcal{V}_{1jkl}) = 0.2 \abs{\mathcal{V}_{1jkl}}$, other values have no uncertainty. Gaussian quadrature points and weights were used both for momenta and angles. For momenta in \eqref{p}, $48$ points were used in the interval $(0 \, \text{fm}^{-1} , 20 \, \text{fm}^{-1})$. For $x$ in \eqref{x}, $48$ points were used in the interval $(-1 , 1)$. All discrete quantum numbers with total angular momentum 
	less than or equal to $5$ were used in the partial wave representation. 
	\ra{The error bars show the standard deviation calculated obtained using backpropagation for a given energy.}}
	\label{f1}
\end{figure*}

Equation \eqref{de1} is solved by scanning different values of $E$ and searching for an energy for which
$\lambda \approx 1$. This is illustrated in Figure \ref{f1} where we limit the errors to a single scalar function
($\alpha(1) = 0.2$, $\alpha(j) = 0$ if $j \ne 1$). 
The dots mark the eigenvalue that is
closest to $1$ for a given energy. The error bars are calculated from Eq. \eqref{err} with the gradients obtained from backpropagation. 
We will approximate this data using a linear function, with the solid line being the
result of a least squares regression.
The plot title contains an estimate of the deuteron binding energy $E_{d}$, calculated as the energy value for which the line crosses $1$ and the error associated with this result. 

\begin{figure}[h]
	\begin{center}
		\includegraphics[width = 0.9 \linewidth]{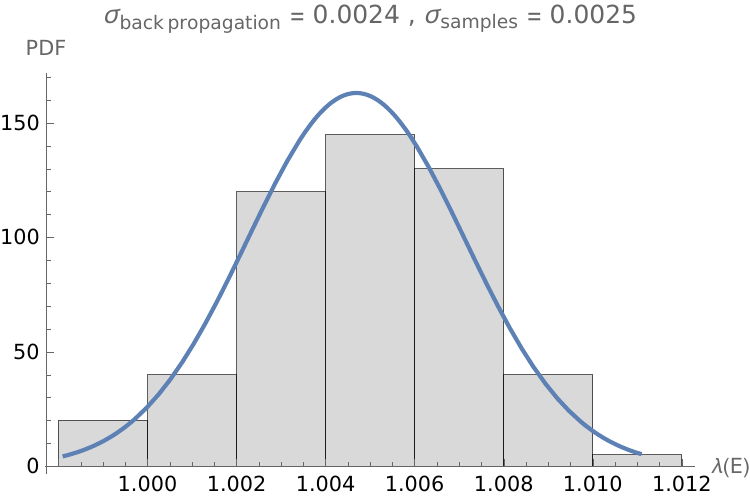}
	\end{center}
	\caption{Histogram of eigenvalues closest to $1$ for single energy $E = -2.14206$ MeV from Figure \ref{f1}. The blue line
	is the Normal distribution Probability Density Function (PDF) with parameters calculated using \eqref{err}. Gaussian quadrature points and weights were used both for momenta and angles. For momenta in \eqref{p}, $48$ points were used in the interval $(0 \, \text{fm}^{-1} , 10 \, \text{fm}^{-1})$. For $x$ in \eqref{x}, $48$ points were used in the interval $(-1 , 1)$. All discrete quantum numbers with total angular momentum 
	less than or equal to $5$ were used in the partial wave representation.
	The data was tested for normality using the Kolmogorov-Smirnov test, the p-value is $0.955$. 
	}
	\label{f1a}
\end{figure}

For one value of the energy in Figure \ref{f1}, there are additional gray dots. These values were 
calculated by directly sampling new $\mathcal{V}'$ ``tensors'' from the normal distribution with standard deviation \eqref{paramerrors} 
and mean $\mathcal{V}$,
solving \eqref{de1} and using the eigenvalue that is closest to $1$. A histogram of these
sampled points is compared with the uncertainty calculated using backpropagation in Figure \ref{f1a}. The blue solid line
is the normal distribution probability density function with parameters determined from \eqref{err}. Although the error
$\alpha(1)$ is quite large a comparison with the histogram shows that the sampled distribution is well approximated by a Gaussian. 
The values of the standard deviations are also in good agreement for backpropagation and sampling. 
Finally, Figure \ref{f1} contains an estimation of the deuteron binding energy and the uncertainty of this estimate. These values were obtained in two steps. Firstly,
the uncertainty for eigenvalues at individual energies,
obtained from \eqref{err} with $\frac{\partial{\lambda}}{\partial{\mathcal{V}_{ijkl}}}$ calculated using backpropagation and assuming \eqref{paramerrors}, is taken into account when calculating the least squares linear fit.
Secondly, the deuteron binding energy is calculated as the intersection of the fitted line with $1$ and the uncertainty of the deuteron binding energy is calculated from the errors of the least squares regression.
Note that the bound state calculations were performed considering all partial wave states with total angular momentum
less than or equal $5$. For the deuteron it is sufficient to consider only two partial wave states, so increasing the number of considered states served as a benchmarking test for our method. Indeed, including this large number of partial
waves still reproduces the correct binding energy for the Bonn-B \cite{Machleidt1989} potential.

\begin{figure*}[h]
	\begin{center}
		\includegraphics[width = 0.75 \linewidth]{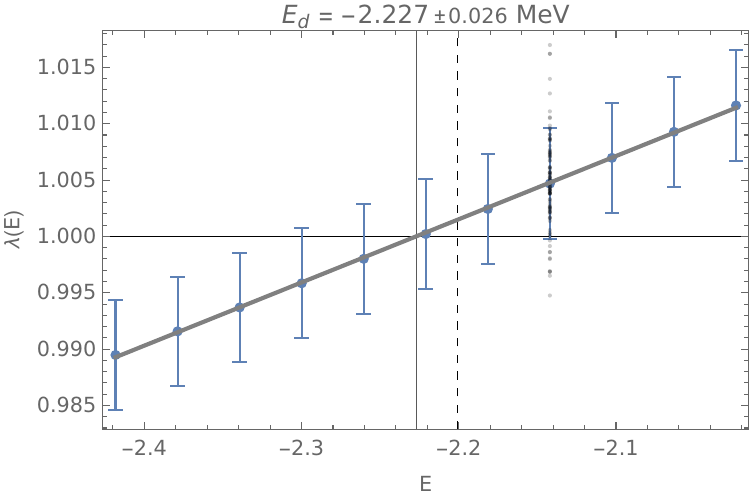}
	\end{center}
	\caption{Similar to Figure \ref{f1}, eigenvalues closest to $1$ for different
	values of $E$ in MeV from \eqref{de1}. 
	The standard deviation of the scalar function values is \ra{taken to be} $\sigma(\mathcal{V}_{1jkl}) = 0.4 \abs{\mathcal{V}_{1jkl}}$, other values have no uncertainty.}
	\label{f3}
\end{figure*}

\begin{figure}[h]
	\begin{center}
		\includegraphics[width = 0.9 \linewidth]{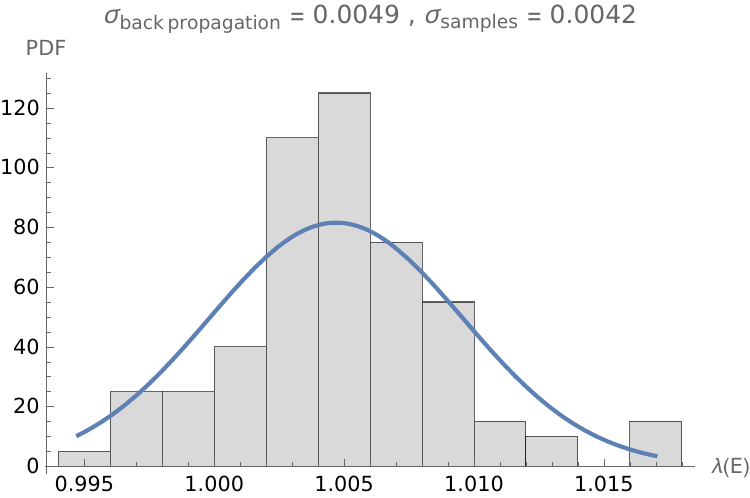}
	\end{center}
	\caption{Similar to Figure \ref{f1a}, histogram of eigenvalues closest to $1$ for single energy $E = -2.14206$ MeV from Figure \ref{f3}. The blue line
	is the Normal distribution Probability Density Function (PDF) with parameters calculated using \eqref{err}.
	The data was tested for normality using the Kolmogorov-Smirnov test, the p-value is $0.382$. 
	}
	\label{f3a}
\end{figure}

Using gradients, a linear approximation of the $\lambda$ dependence on the scalar function values
can be constructed allowing the errors to be calculated using simple methods. 
This approximation is justified 
if the uncertainties of the scalar functions result
in small deviations from the mean.
It is interesting to investigate a situation where
this approximation is no longer valid.
Figures \ref{f3} and \ref{f3a} contain results where the relative uncertainty $\alpha$ is increased.
This results in an increased uncertainty of determining the deuteron binding energy.
Additionally, the values
of the standard deviations are in slightly worse agreement as can be seen in Figure \ref{f3a}.

\begin{figure*}[h]
	\begin{center}
		\includegraphics[width = 0.75 \linewidth]{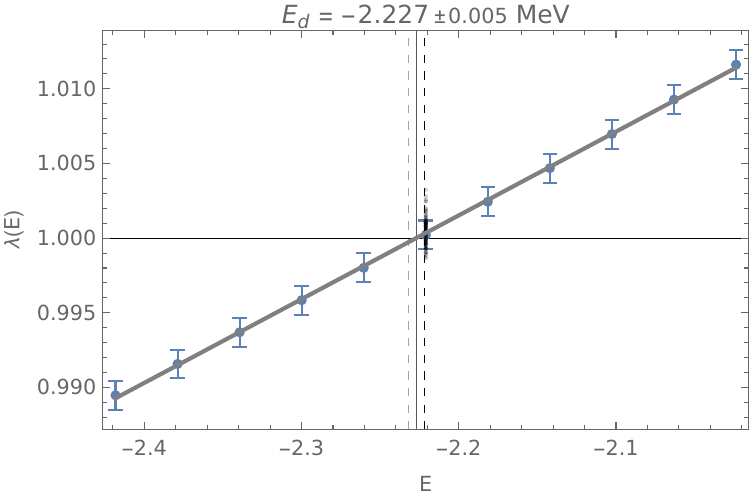}
	\end{center}
	\caption{Similar to Figure \ref{f1}, eigenvalues closest to $1$ for different
	values of $E$ in MeV from \eqref{de1}. 
	The standard deviation of the scalar function values is \ra{taken to be} $\sigma(\mathcal{V}_{ijkl}) = 0.05 \abs{\mathcal{V}_{ijkl}}$, for all $i$, $j$, $k$, and $l$.
	}
	\label{f2}
\end{figure*}

\begin{figure}[h]
	\begin{center}
		\includegraphics[width = 0.9 \linewidth]{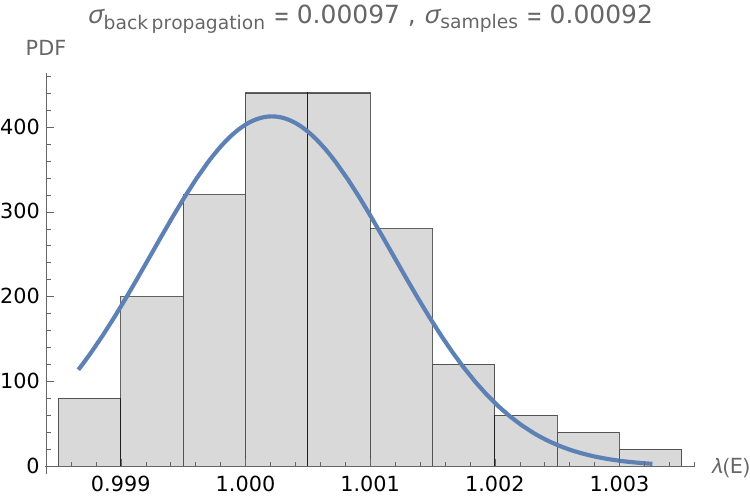}
	\end{center}
	\caption{Similar to Figure \ref{f1a}, histogram of eigenvalues closest to $1$ for single energy $E = -2.22099$ MeV from Figure \ref{f2}. The blue line
	is the Normal distribution Probability Density Function (PDF) with parameters calculated using \eqref{err}.
	The data was tested for normality using the Kolmogorov-Smirnov test, the p-value is $0.424$.
	}
	\label{f2a}
\end{figure}

Figures \ref{f2} and \ref{f2a} show results where we are not limiting the uncertainty to 
a single scalar function. The relative uncertainty $\alpha = 0.05$ is small compared to the previous examples
and this results in a small uncertainty for the deuteron binding energy. The sampled eigenvalues are in good agreement
with the normal distribution whose parameters were obtained using backpropagation as can be seen in Figure \ref{f2a}.

\begin{figure}[h]
	\begin{center}
		\includegraphics[width = 0.9 \linewidth]{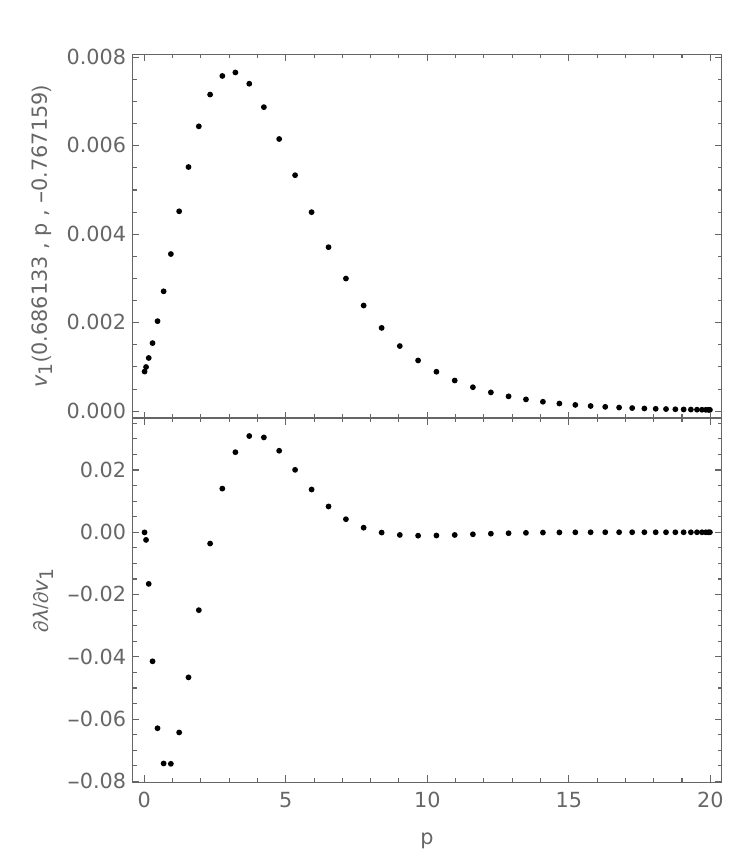}
	\end{center}
	\caption{Upper plot: selected values of $v_{1}(0.686133 , p , -0.767159)$ from \eqref{com} are plotted for different
	values of $p$ in $\text{fm}^{-1}$. Lower plot: the gradient of the eigenvalue $\lambda$ from \eqref{de1} that
	is closest to $1$ ($\lambda = 1.000212$)
	for energy $-2.22099$ MeV with respect
	to the corresponding scalar function values from the upper plot.
	}
	\label{ff}
\end{figure}

Finally, Figure \ref{ff} shows the gradient of the eigenvalue closest to $1$ from \eqref{de1} 
for selected scalar function values and error estimates for different values of the relative error 
are 
gathered in Table \ref{tabledeuteron}. 
The first column of this table contains the relative error $\alpha$. It was assumed that this value is the
same for all scalar functions. The second and third column contain the calculated deuteron binding energy and the
error associated with this estimate. For all considered values of the relative error, the calculated deuteron binding
energy is similar and the error increases in a roughly linear fashion with $\alpha$.

\begin{table}[h]
\centering
\begin{tabular}{|c|c|c|}
\hline
	$\alpha(i = 1 \ldots 6)$ & $E_{d}$ & $\sigma(E_{d})$ \\
	\hline
 0.01 & -2.2272 & 0.0010 \\
 0.03 & -2.2272 & 0.0031 \\
 0.05 & -2.2272 & 0.0052 \\
 0.07 & -2.2272 & 0.0073 \\
 0.1 & -2.2272 & 0.0104 \\
\hline
\end{tabular}
	\caption{The deuteron binding energy and its uncertainty
	for different values of the relative error $\alpha$. 
	The second column contains roughly the same value for
	all relative errors - the deuteron binding energy in MeV. 
	The third column contains errors of the energy estimate in MeV obtained using backpropagation.
	The relationship
	between the deuteron binding energy error value and the relative error is roughly linear. 
	}
\label{tabledeuteron}
\end{table}

\section{Nucleon - nucleon scattering}
\label{nnintroduction}

In order to calculate the nucleon - nucleon (NN) scattering observables we will solve
the Lippmann - Schwinger equation:
\begin{equation}
	t(z) = V + V G_{0}(z) t(z).
	\label{lse}
\end{equation}
to obtain the ``t matrix'', $t(z)$, for the complex argument
\[
	z = E + i \epsilon
\]
where $E > 0$ is the energy of the 2N system. The infinitesimal $\epsilon$ is real and we take the limit $\epsilon \rightarrow 0$.
In equation \eqref{lse}, the free propagator $G_{0}$ has the form:
\[
	G_{0}(z) \equiv (z - H_{0})^{-1}
\]
where $H_{0} = \frac{\vb{p}^{2}}{m}$ is the free Hamiltonian operator with $\vb{p}$ being the relative
NN momentum and $m$ being the nucleon mass. 
With the assumption that the potential satisfies \eqref{2npot} 
equation \eqref{lse} separates and can be solved independently for each isospin case $t^{t m_{t}}(z)$.

In order to arrive at a form of \eqref{lse} that can be used in 
numerical calculations, the identity operator can be inserted
between operators. For partial wave states \eqref{pwstates} it is given by:
\begin{equation} 
 1 = \int_{0}^{\infty} \text{d}p  \, p^{2} \sum_{\alpha} \dyad{p \alpha}{p \alpha}
	\label{id}
\end{equation}
with $\ket{p \alpha}$ being partial wave states \eqref{pwstates} and $\alpha$
being the complete sets of discrete quantum numbers. In practice the sum over $\alpha$ is
limited to a finite number of 2N states and the integral upper limit is replaced with a 
finite cut-off, turning \eqref{id} into an approximation. In the first step, using \eqref{com} and inserting \eqref{id}
between the operators, the Lippmann - Schwinger equation \eqref{lse} can be transformed into:
\begin{align} \nonumber
	\braket{p' \alpha'}{\phi} - \int_{0}^{\infty} \text{d}k k^{2} \frac{1}{p_{0}^{2} - k^{2} + i \epsilon} \\
	\sum_{\beta} m \mel{p' \alpha'}{V^{t m_{t}}}{k \beta} \braket{k \beta}{\phi} = \braket{p' \alpha'}{\nu}
	\label{lse1}
\end{align}
where 
\begin{equation}
	\ket{\phi} \equiv t^{t m_{t}}(E + i \epsilon = \frac{p_{0}^{2}}{m} + i \epsilon) \ket{p \alpha},
	\label{phi}
\end{equation}
and
\begin{equation}
	\ket{\nu} \equiv V^{t m_{t}} \ket{p \alpha}.
	\label{nu}
\end{equation}
Equation \eqref{lse1} can be solved separately for the states $\ket{\phi}$ with given isospin ($t, m_{t}$), energy 
($E = \frac{p_{0}^{2}}{m}$), initial relative momentum ($p$) and discrete quantum numbers ($\alpha$).
The integral in \eqref{lse1} can be evaluated using the standard approach \cite{Golak2009} by replacing infinity 
with a large finite momentum cut-off
$\bar{p}$ and taking the limit $\epsilon \rightarrow 0$ to obtain:
\begin{align} \nonumber
	\braket{p' \alpha'}{\phi} - \int_{0}^{\bar{p}} \text{d}k \frac{1}{p_{0}^{2} - k^{2}} \\ \nonumber
	\left( \sum_{\beta} k^{2} m \mel{p' \alpha'}{V^{t m_{t}}}{k \beta} \braket{k \beta}{\phi} - \right. \\ \nonumber
	\left. \sum_{\beta} p_{0}^{2} m \mel{p' \alpha'}{V^{t m_{t}}}{p_{0} \beta} \braket{p_{0} \beta}{\phi} \right) - 
	\frac{1}{2} p_{0} \sum_{\beta} m \\  
	\mel{p' \alpha'}{V^{t m_{t}}}{p_{0} \beta} \braket{p_{0} \beta}{\phi}  
	\left( \ln \frac{\bar{p} + p_{0}}{\bar{p} - p_{0}} - i \pi  \right) = 
	\braket{p' \alpha'}{\nu}
	\label{lse2}
\end{align}
The implementation of the matrix form of the linear equation \eqref{lse2} for state $\ket{\phi}$ using the \textbf{pytorch} 
library is discussed
in subsection \ref{nntorch}.

\subsection{Implementation using \textbf{pytorch}}
\label{nntorch}

In order to turn \eqref{lse2} into a matrix equation that
can be solved using the \textbf{pytorch} library it is useful to extend
the set of discrete values of the relative 2N momentum \eqref{p}:
\begin{equation}
	p_{1} , p_{2} \ldots p_{N} , p_{N + 1} \equiv p_{0} \label{p1}
\end{equation}
and to set the integration weight $\mathcal{W}^{p}_{N + 1} \equiv 0$.
With this extended set of momentum values and the $\mathcal{V}^{pwd}$ ``tensor'' from \eqref{vpwd}
containing the partial wave matrix elements of the 2N potential equation \eqref{lse2}
becomes:
\begin{equation}
	\sum_{k = 1}^{N + 1} \sum_{\beta} \mathcal{A}^{t m_{t}}(p_{0})_{j \gamma k \beta} \mathcal{\phi}_{k \beta} = \mathcal{\nu}_{j \gamma} ,
\end{equation}
where the second sum is over a finite set of considered discrete quantum numbers $\beta$.
The states \eqref{phi} and \eqref{nu} are represented by \textbf{pytorch} ``tensors'' 
$\mathcal{\phi}_{k \beta} \equiv \braket{p_{k} \beta}{\phi}$ and $\mathcal{\nu}_{j \gamma} \equiv \braket{p_{j} \gamma}{\nu}$
respectively.
The explicit form of the elements of $\mathcal{A}^{t m_{t}}$ is more complicated
than \eqref{kdeuteron} but it can be directly read off from \eqref{lse2}.
The practical implementation of $\mathcal{A}^{t m_{t}}(p_{0})_{j \gamma k \beta}$ is made easier
by utilizing the capabilities of \textbf{pytorch} that allow to automatically broadcast functions
over ``tensors'' and 
by treating the case $k = N + 1$ separately.

In order to calculate observables $p$ is set to $p_{0}$ in \eqref{phi} and \eqref{nu}. Finally, all ``tensors'' are
appropriately reshaped before using the \textbf{torch.linalg.solve} function to solve the linear equation
for $\ket{\phi}$. 
Since all operations are performed within the \textbf{pytorch} library, gradients of the observables
with respect to the scalar functions that define the interaction \eqref{com} are available allowing the estimation
of errors. Similarly as in the deuteron case, we will assume that the uncertainty of the scalar functions
that define the 2N interaction has the form \eqref{paramerrors} and assume that there are
no correlations so that the errors can be calculated using a formula similar to \eqref{err} but with $\lambda$
replaced by the relevant observable.
In the next 
subsection \ref{nnresults} we will present result for phase-shifts calculated for different partial wave
channels and energies. 

\subsection{Results for nucleon - nucleon scattering}
\label{nnresults}

{To demonstrate our method we computed phase shifts for two uncoupled channels $^{1}S_{0}$ and $^{1}P_{1}$ and for the two coupled cases: $^{3}S_{1} - {^3D}_{1}$ and $^{3}P_{2} - {^{3}F_{2}}$}
using the Bonn-B \cite{Machleidt1989} potential.
Calculations {at $p_0=2$~fm$^{-1}$ 
and $\alpha(i = 1, \ldots, 6)=0.1$.}
for $^{1}S_{0}$ resulted in a value of $-11.8988$ degrees. The error estimated associated with this
result was calculated using backpropagation and its value, $3.38991$ degrees, is in agreement with 
the standard deviation calculated using direct sampling of scalar values that define the potential.
The sampling procedure is the same as in the deuteron case. The calculation is repeated for each of
the $100$ samples that contain new scalar function values resulting in $100$ phase shift values. The standard
deviation of these values is $3.46838$.

Results for the $^{1}P_{1}$ channel are gathered in Tables \ref{tab1} and \ref{tab2}. Table \ref{tab1} 
contains phase shifts calculated with $p_{0} = 2 \, \text{fm}^{-1}$
for different values of the relative error $\alpha$. The relative error was assumed to be the same
for all scalar functions that define the potential. The second and third column contain estimates
of the error associated with the phase shift in column two. These columns were calculated using
backpropagation and direct sampling, respectively. The error estimates are in agreement.
Table \ref{tab2} contains phase shifts calculated for different values of $p_{0}$. 
The second column contains the calculated 
value in degrees. The third and fourth columns contain errors estimated using backpropagation and direct sampling in degrees.
Note the large error and discrepancy between the third and fourth column for $p_{0} = 0.5 \, \text{fm}^{-1}$ and $p_{0} = 4.0 \, \text{fm}^{-1}$.
This can be explained by the close proximity of these values to momentum values 
$0.4999 \, \text{fm}^{-1}$ and $4.0200 \, \text{fm}^{-1}$ 
from the Gaussian quadrature points used in calculating integrals and demonstrates the sensitivity
of the approach to the details of the numerical implementation. 
The large uncertainty obtained using backpropagation can be used as a signal
indicating the necessity to reconsider the parameters of the numerical calculation. This is especially visible for $p_{0} = 4 \, \text{fm}^{-1}$ where it
is evident that the choice of integration points effects on the 
uncertainty calculated directly from sampling.

In addition to results for the two uncoupled channels that were calculated using only a single partial wave state, 
calculations for two coupled channels $^{3}S_{1} - {^{3}D_{1}}$ and $^{3}P_{2} - {^{3}F_{2}}$ 
were carried out {at $p_0=2$~fm$^{-1}$} and required using two partial waves.
The results for $^{3}S_{1} - {^{3}D_{1}}$ are gathered in Table \ref{tab3} and results for $^{3}P_{2} - {^{3}F_{2}}$
are gathered in Table \ref{tab4}. 
The error estimates obtained using
backpropagation are in agreement with estimates obtained using direct sampling with $100$ samples. {Of course their absolute values depend on the arbitrarily assumed $\alpha$ thus they do not carry any information on quality of the specific potential.}

\begin{table}[h]
\centering
\begin{tabular}{|c|c|c|c|}
\hline
	$\alpha(i = 1 \ldots 6)$ & $s$ & $\sigma^{\text{\begin{tiny}backpropagation\end{tiny}}}(s)$ & $\sigma^{\text{\begin{tiny}samples\end{tiny}}}(s)$ \\
\hline
 0.01 & -30.1 &  0.5 &  0.5 \\
 0.03 & -30.1 &  1.4 &  1.4 \\
 0.05 & -30.1 &  2.4 &  2.7 \\
 0.07 & -30.1 &  3.3 &  3.2 \\
 0.1 & -30.1 &  4.8 &  4.6 \\
\hline
\end{tabular}
	\caption{$^{1}P_{1}$ phase shifts calculated for nucleon - nucleon scattering scattering with $p_{0} = 2 \, \text{fm}^{-1}$
	for different values of the relative error $\alpha$. The second column contains the same value for
	all relative errors - the calculated phase shift 
	value in degrees. The third and fourth column contains errors estimated using backpropagation and direct sampling in degrees, respectively. The relationship
	between the phase shift error value and the relative error is roughly linear.
	}
	\label{tab1}
\end{table}

\begin{table}[h]
\centering
\begin{tabular}{|c|c|c|c|}
\hline
	$p_{0}$ & $s$ & $\sigma^{\text{\begin{tiny}backpropagation\end{tiny}}}(s)$ & $\sigma^{\text{\begin{tiny}samples\end{tiny}}}(s)$ \\
\hline
 0.5 & -6.3 & 3225.6 &  0.4 \\
 1. & -14.8 &  0.4 &  0.4 \\
 2. & -30.1 &  2.4 &  2.7 \\
 3. & -39.8 &  1.2 &  1.1 \\
 4. & -41.9 &  217.8 &  26.0 \\
 5. & -43.0 &  1.7 &  1.6 \\
 6. & -47.9 &  2.2 &  2.0 \\
 7. & -55.5 &  3.5 &  3.5 \\
\hline
\end{tabular}
	\caption{$^{1}P_{1}$ phase shifts calculated for nucleon - nucleon scattering scattering with 
	the relative error $\alpha(i = 1 \ldots 6) = 0.05$ for different values of $p_{0}$ in $\text{fm}^{-1}$ in the first
	column.
	The second column contains the calculated 
	value in degrees. The third and fourth column contains errors estimated using backpropagation and direct sampling in degrees, respectively.
	}
	\label{tab2}
\end{table}

\begin{table}[h]
\centering
\begin{tabular}{|c|c|c|c|}
\hline
	quantity & value & $\sigma^{\text{\begin{tiny}backpropagation\end{tiny}}}(\text{\begin{tiny}quantity\end{tiny}})$ & $\sigma^{\text{\begin{tiny}samples\end{tiny}}}(\text{\begin{tiny}quantity\end{tiny}})$ \\
\hline
	$\delta_{M}$ & 0.2 &  1.4 &  1.4 \\
	$\delta_{S}$ & -24.3 &  0.4 &  0.4 \\
	$\epsilon$ & -4.3 &  0.5 &  0.5 \\
\hline
\end{tabular}
	\caption{Phase shift and mixing parameters for the coupled $^{3}S_{1} - ^{3}D_{1}$ channel. The calculations were performed with $p_{0} = 2\,\text{fm}^{-1}$ and assumed 
	the relative error $\alpha(i = 1 \ldots 6) = 0.1$. The third and fourth column contains error estimates
	obtained using backpropagation and direct sampling using $100$ samples, respectively.
	}
\label{tab3}
\end{table}

\begin{table}[h]
\centering
\begin{tabular}{|c|c|c|c|}
\hline
	quantity & value & $\sigma^{\text{\begin{tiny}backpropagation\end{tiny}}}(\text{\begin{tiny}quantity\end{tiny}})$ & $\sigma^{\text{\begin{tiny}samples\end{tiny}}}(\text{\begin{tiny}quantity\end{tiny}})$ \\
\hline
	$\delta_{M}$ & 17.23 &  0.49 &  0.50 \\
	$\delta_{S}$ & 0.15 &  0.24 &  0.25 \\
	$\epsilon$ & 1.72 &  0.07 &  0.07 \\
\hline
\end{tabular}
	\caption{Phase shift and mixing parameters for the coupled $^{3}P_{2} - ^{3}F_{2}$  channel. The calculations were performed with $p_{0} = 2\,\text{fm}^{-1}$ and assumed 
	the relative error $\alpha(i = 1 \ldots 6) = 0.05$. The third and fourth column contains error estimates
	obtained using backpropagation and direct sampling using $100$ samples, respectively.}
\label{tab4}
\end{table}

\section{Summary and outlook}
\label{so}

In this paper we show that software libraries designed for machine learning can be used in practical
few-nucleon quantum mechanical calculations. One benefit of using these libraries, is the possibility
to use backpropagation to estimate how the errors propagate from the model parameters, in our case
the values of the scalar functions that determine the two-nucleon potential, to observables. {In our calculations we use the popular {\bf Python pytorch} library~\cite{NEURIPS2019_9015}. It has a straightforward interface that allows to easily extend its use beyond machine learning applications and utilize the built in implementation of backpropagation.}  

To demonstrate this approach we associate an uncertainty with the scalar function values and
investigate the effect of this uncertainty on the error of determining the deuteron binding energy
and nucleon - nucleon scattering phase shifts.
The assumed uncertainties were chosen in order to demonstrate
the validity of the approach, for more practical results the numerical errors and truncation errors from higher
orders of the chiral expansion could be considered. {
The actual parameter uncertainty values for the SMS chiral potential~\cite{Reinert_SMS} are typically or very small ($\ll 1$\%) or up to 3\%, and only in few cases are above 10\%. For some of them the correlation is non-negligible.}

Currently we assume that the errors of the scalar function values are uncorrelated. In future work
it will be interesting to see the effect of correlated errors on the deuteron binding energy estimate
or other observables. This would open up the possibility to use more realistic error estimates in the
calculation. Furthermore, we plan to extend our calculations to the three-nucleon system to investigate
the effects of theoretical and numerical errors on observables obtained using the new generation of 
few-nucleon forces. {We believe that because the method presented here is suitable for calculations based on solving an eigenequation or a system of linear equations, it can be also used in nuclear structure calculations. Even if using the {\bf Python pytorch} library directly may be impractical in that case, the analogous code can be developed in standard programming languages.}

\section*{Acknowledgements}
{This work was supported by the National Science Centre, Poland under Grant \newline IMPRESS-U 2024/06/Y/ST2/00135.
It was also supported in part by
the Excellence Initiative – Research University Program at the Jagiellonian University in
Krak\'ow. The numerical calculations were partly performed on the supercomputers of the
JSC, J\"ulich, Germany.}

\section*{Code and data availability statement}

The code and plot data from the current study are available as a GitHub repository at \newline
\begin{scriptsize}
\url{https://github.com/kacpertopol/uncertainties_backpropagation}
\end{scriptsize}.

\bibliographystyle{ieeetr}
\bibliography{main}

\end{document}